\documentclass[twocolumn,aip,jcp,showpacs,preprint,amsmath,amssymb,showkeys,citeautoscript,10pt]{revtex4-1}    

\usepackage[utf8]{inputenc}
\usepackage[TS1,OT1,T1]{fontenc}
\usepackage[english]{babel}
\usepackage{lmodern} 

\usepackage{multirow}

\usepackage{array}
\usepackage{amsmath}
\usepackage{amssymb}
\usepackage{amsthm}
\usepackage{dsfont}
\RequirePackage{pgfplots}  
\usepackage{epstopdf}
\usepackage{epsfig}
\usepackage{placeins}
\usepackage{setspace}
\usepackage{subfig}
\usepackage{xfrac}
\usepackage{url}

\usepackage[normalem]{ulem}
\newcommand{\stkout}[1]{\ifmmode\text{\sout{\ensuremath{#1}}}\else\sout{#1}\fi}

\newcommand\R{\mathbb{R}}

\newcommand\Rc{R_{\mathrm{cut}}}

\newcommand\e{\mathrm{e}}

\newcommand{\angstrom}{\mbox{\normalfont\AA}}

\graphicspath{{img/}}

\begin{document}

\title{Learning molecular energies using localized graph kernels}

\author{Gr\'egoire Ferr\'e}
\affiliation{Universit\'e Paris-Est, CERMICS (ENPC), F-77455 Marne-la-Vall\'ee, France}
\affiliation{Theoretical Division and CNLS, Los Alamos National Laboratory, Los Alamos, NM 87545, USA}
\author{Terry Haut}
\affiliation{Computational Physics and Methods Group (CCS-2), Los Alamos National Laboratory, Los Alamos, NM, 87545, USA}
\author{Kipton Barros}
\affiliation{Theoretical Division and CNLS, Los Alamos National Laboratory, Los Alamos, NM 87545, USA}

\date{\today}

\begin{abstract}
Recent machine learning methods make it possible to model potential energy of atomic
configurations with chemical-level accuracy (as calculated from ab-initio calculations) 
and at speeds suitable for molecular dynamics simulation.
Best performance is achieved when the known physical constraints are encoded in the machine 
learning models. For example, the atomic energy is invariant under global translations 
and rotations; it is also invariant to permutations of same-species atoms. Although simple 
to state, these symmetries are complicated to encode into machine learning algorithms.
In this paper, we present a machine learning approach based on graph theory
that naturally incorporates translation, rotation, and permutation symmetries. Specifically, 
we use a \emph{random walk graph kernel} to measure the similarity of two adjacency matrices, 
each of which represents a local atomic environment.
This Graph Approximated Energy (GRAPE) approach is flexible and admits many possible extensions.
We benchmark a simple version of GRAPE by predicting atomization energies on a standard dataset of organic molecules.
\end{abstract}

\maketitle


\section{Introduction}


There is a large and growing interest in applying machine 
learning methods to infer molecular properties and
potential energies~\cite{Ho1996,Bettens1999,Braams2009,Molecular2013,rupp2015machine,bartok2015gaussian}
based upon data generated from \emph{ab initio} quantum calculations.
Steady improvement in the accuracy, transferability, and computational efficiency of the 
machine learned models has been achieved. A crucial enabler, and ongoing challenge, is 
incorporating prior physical knowledge into the methodology. In particular, we know that 
potential energies should be invariant with respect to translations and rotations of atomic positions, 
and also with respect to permutations of identical atoms.
The search for new techniques that naturally encode physical invariances is of continuing interest.

A common modeling approach is to decompose the energy of an atomic configuration as a sum 
of local contributions, and to represent each local atomic environment in terms of 
 descriptors (\emph{i.e.} features) that are inherently rotation and permutation 
invariant~\cite{behler2011atom, Bartok2013repr}.
Such descriptors may be employed in, e.g., linear regression~\cite{Thompson2015} or neural 
network models~\cite{behler2007,handley2014next,behler2014ref,behler2015constructing,Schutt2017}.
Recently, sophisticated descriptors have been proposed based on an expansion of invariant 
polynomials~\cite{shapeev2015moment} and, inspired by convolutional neural networks, on a 
cascade of multiscale wavelet transformations~\cite{hirn2015quantum,hirn2016wavelet}.

Kernel learning is an alternative approach that can avoid direct use of 
descriptors~\cite{smola1998learning,scholkopf2001learning}. Instead, one specifies a kernel 
that measures similarity between two inputs and, ideally, directly satisfies known invariances.
An advantage of kernel methods is that the accuracy can systematically improve by adding
new configurations to the dataset. A disadvantage is that the computational 
cost to build a kernel regression model typically scales cubically with dataset size. Kernel methods 
that model potential energies include Gaussian Approximation 
Potentials (GAP)~\cite{Bartok2010GAP}, Smooth Overlap of Atomic 
Potentials (SOAP)~\cite{Bartok2013repr,bartok2015gaussian}, methods based upon eigenvalues of the Coulomb matrix~\cite{Rupp2012}, Bag-of-Bonds~\cite{Hansen2015}, 
and others~\cite{Molecular2013,schutt2014, li2015molecular,ferre2015permutation}. However, encoding the physical invariances may impose challenges. For example, to achieve rotational invariance in GAP and SOAP, one 
requires explicit integration over the relative rotation of two configurations. To achieve permutation invariance in 
methods based upon the Coulomb matrix, one commonly restricts attention to sorted descriptors (e.g. matrix eigenvalues or Bag-of-Bonds),
thus introducing singularities in the regression model, and potentially also sacrificing discriminative power.

Here, we present the Graph Approximated Energy (GRAPE) framework.
GRAPE is based on two central ideas: we represent local atomic environments as 
weighted graphs based upon pairwise distances, for which rotation invariance is inherent, and 
we compare local atomic environments by leveraging well known graph kernel methods
invariant with respect to node permutation.
There is much flexibility in implementing these ideas.
In this paper, we define the weighted adjacency matrix elements (\emph{i.e.} the edge weights) 
through an analogy with SOAP~\cite{Bartok2013repr}, and we use generalized random walk graph 
kernels~\cite{gartner2002exponential,gartner2003graph,vishwanathan2010graph} to define 
similarity between graphs. Graph kernels have commonly been employed to compare molecules and predict their properties~\cite{drews2000drug,smola2003kernels,borgwardt2005shortest,ralaivola2005graph,sharan2006modeling,schietgat2008efficiently,smalter2009graph,vishwanathan2010graph,gauzere2012two}, 
but it appears that prior applications to energy regression are limited~\cite{sunthesis}. 
We demonstrate with benchmarks that GRAPE shows promise for energy regression tasks.

The rest of the paper is organized as follows. Section~\ref{sec:review} reviews 
necessary background techniques: kernel ridge regression of energies, the SOAP kernel, and 
random walk graph kernels. We merge these tools to obtain our GRAPE kernel, presented 
in Sec.~\ref{sec:grape}. Finally, we demonstrate the performance of GRAPE on a standard 
molecular database in Sec.~\ref{sec:application} and summarize our results in 
Sec.~\ref{sec:discussion}.


\section{Review of relevant methods}
\label{sec:review}

\subsection{Kernel modeling of potential energy}
\label{sec:regression}

Here we review the machine learning technique of kernel ridge regression
and its application to modeling potential energy landscapes. We assume
a dataset $(x_{i},y_{i})_{i=1\dots N}$, where $x_{i}$ represents
an atomic configuration and $y_{i}=E(x_{i})$ its corresponding energy.
The goal is to build a statistical approximation $\hat{E}$ that accurately
estimates the energy of new configurations.

A common approach begins with \emph{descriptors} (also called features
or order parameters) $(h_{m})_{m=1\dots M}$, designed
to capture various aspects of a configuration $x$. Often $h_{m}(x)$ describes the
geometry of a local atomic environment. Many descriptors suitable
for modeling potential energy have been developed~\cite{Bartok2013repr}.
The descriptors can be used, for example, with simple linear regression~\cite{Thompson2015},
\begin{equation}
\hat{E}(x)=\sum_{m=1}^{M}\beta_{m}h_{m}(x),\label{eq:linear_reg}
\end{equation}
with learned regression coefficients $\beta_{m}$, or as inputs to a more complicated model such as a neural network~\cite{behler2015constructing}.

Kernel methods are an alternative machine learning approach, and avoid direct use of
descriptors. Instead, the starting point is a kernel $K(x,x')$ that measures the similarity between configurations $x$ and $x'$. We use
kernel ridge regression, for which the approximated energy reads
\begin{equation}
\label{eq:krr}
\left\{
\renewcommand{\arraystretch}{2} 
\begin{array}{rl}
\displaystyle \hat{E}(x) &\displaystyle =\sum_{i=1}^{N}\hat{\alpha}_{i}K(x_i,x),\\
\displaystyle \hat{\alpha} &\displaystyle =(K+\lambda I)^{-1} y,
\end{array}
\right.
\end{equation}
where $I \in \R^{N\times N}$ is the identity matrix, $K\in \R^{N\times N}$ denotes a matrix with elements
$K(x_{i},x_{j})$, and $y=(y_{1},y_{2},\dots ,y_{N})$ are the energies in the 
dataset. The parameter $\lambda > 0$ serves to regularize the model, and prevents overfitting
by penalizing high frequencies~\cite{smola1998learning}.

Equations~\eqref{eq:krr} can be readily derived
from the linear model,~\eqref{eq:linear_reg},
using least squares error minimization with a ridge penalty; see 
Appendix~\ref{sec:kernelapprox} for details.
The kernel becomes
$K(x,x')=\sum_{m=1}^M h_{m}(x)h_{m}(x')$, and thus satisfies the criteria of symmetry and positive 
semi-definiteness. Conversely,
any function $K$ that is symmetric and positive semi-definite can
be decomposed as above, where $M$ may be infinite~\cite{Mercer1909}.
Thus, $(h_{m})_{m=1 \dots M}$ and $K$ are formally interchangeable.
An advantage of specifying the kernel directly is that the corresponding, potentially
infinite set of descriptors may remain fully implicit.

In the context of modeling potential energies and forces, physical locality is often 
a good approximation~\cite{bartok2015gaussian}. We assume that the energy of an atomic 
configuration $x_i$ with $n_i$ atoms can be decomposed as a sum of local contributions, 
\begin{equation}
E(x)=\sum_{\ell=1}^{n_i}\varepsilon(\tilde{x}_{i,\ell}),\label{eq:locality}
\end{equation}
where $\tilde{x}_{i,\ell}$ denotes a local view of environment $x_i$
centered on atom $\ell$. The index $\ell$ iterates through all $n_i$
atoms of the atomic configuration $x_i$, and we denote by 
$Q=\sum_{i=1}^N n_i$ the total number of local environments in the database. If long-range 
interactions exist, \emph{e.g.} induced by electrostatics, they should be treated separately
from~\eqref{eq:locality}.

The key idea is that statistical regression should begin with the local
energy $\varepsilon$, which we formally decompose as in~\eqref{eq:linear_reg}.
Then, after a series of straightforward steps (Appendix~\ref{sec:local}),
we find~\cite{bartok2015gaussian}
\begin{equation}
\label{eq:kernel_identity}
K(x_i,x_{i'})=\sum_{\ell=1}^{n_i}\sum_{\ell'=1}^{n_{i'}}\tilde{K}(\tilde{x}_{i,\ell}
,\tilde{x}'_{i',\ell'}),
\end{equation}
and the localized kernel regression finally reads for a configuration $x$:
\begin{equation}
\label{eq:local_kernel}
\left\{
\renewcommand{\arraystretch}{2} 
\begin{array}{rl}
\displaystyle \hat{E}(x) &= \displaystyle 
\sum_{i=1}^{N} \hat{\alpha}_{i} \sum_{\ell=1}^{n_i} \sum_{\ell'=1}^{n} 
 \tilde{K}(\tilde{x}_{i,\ell},\tilde{x}_{\ell'}),
\\
\displaystyle \hat{\alpha} & = \displaystyle ( L \tilde{K} L^T + \lambda I)^{-1} y,
\end{array}
\right.
\end{equation}
where the entries of the matrix $\tilde{K}\in \R^{Q\times Q}$ are
$\tilde{K}(x_{i,\ell},x_{i',\ell'})$ (\textit{i.e.} the similarity between local
environments $\ell$ and $\ell'$ in configurations $i$ and $i'$), and
$L\in \R^{N\times Q}$ is such that $L_{i,\ell}=1$ if the local
environment $\ell$ belongs to configuration $i$ and $0$ otherwise.

The important conclusion is that the kernel $\tilde{K}$ for \emph{local}
configurations confers a kernel $K$ for total configurations.
This is a mathematical consequence of the energy decomposition, which encodes locality into our statistical model of global energies. Note that the energies $\varepsilon$ of the local atomic configurations $\tilde x_\ell$ need not be explicitly learned. The relatively small size of $\tilde x_\ell$ improves computational efficiency, and generally enhances transferability of the model.

To summarize, Eqs.~\eqref{eq:kernel_identity} and~\eqref{eq:local_kernel}
provide the recipe for kernel ridge regression of energy. The rest of the
paper discusses approaches to designing the local kernel $\tilde{K}$.


\subsection{SOAP kernel}
\label{sec:soap}

Here we review the powerful SOAP method~\cite{Bartok2013repr,bartok2015gaussian} for regression of 
potential energy landscapes. The structure of the SOAP kernel will inspire our design of 
GRAPE, which we present in Sec.~\ref{sec:grape}.

We begin by representing the local configuration $\tilde x_\ell$ with $\tilde{n}_{\ell}$ neighbors
as a smooth atomic density centered on atom $\ell$,
\begin{equation}
\label{eq:local_density}
\rho_{\tilde x_\ell}(\mathbf{r})=\sum_{i=1}^{\tilde{n}_{\ell}}
\omega(z_{i})f_{\mathrm{c}}(|\mathbf{r}_{i\ell}|)\varphi_{\sigma}(|\mathbf{r}-\mathbf{r}_{i\ell}|).
\end{equation}
Here $\mathbf{r}_{i\ell} = \mathbf{r}_{i}-\mathbf{r}_{\ell}$, with $\mathbf{r}_{i}$ the 
position of the $i^{\mathrm{th}}$ atom in $\tilde x_\ell$ and $z_{i}$ its atomic number
with associated weight $\omega(z_i)$.
There is flexibility is choosing the coefficient $\omega(z_i)$, provided that 
it is an injective function of the atomic number $z_i$ for atom $i$.
We also define a Gaussian function to smooth the atomic densities,
\begin{equation}
\label{eq:soapphi} 
\varphi_{\sigma}(r) =\exp(-r^{2}/2 \sigma^2),
\end{equation}
and a cut-off function,
\begin{equation}
\label{eq:soapfc} 
f_{\mathrm{c}}(r) = \left\{
\renewcommand{\arraystretch}{1.5} 
\begin{array}{ll}
\frac{1}{2}\left[1+\cos\left(r\pi / \Rc \right)\right] & \quad r \leq \Rc,\\
0 & \quad r \geq \Rc.
\end{array}
\right. 
\end{equation}
The parameter $\sigma > 0$ sets the smoothing length scale,
and $\Rc > 0$ the cutoff distance.
Atom $i$ contributes to $\rho_{\tilde x_\ell}(\mathbf{r})$ only if sufficiently close to 
atom $\ell$, namely if $|\mathbf{r}_{i\ell}|<\Rc$.
We define $\tilde x_\ell$ to include only those atoms in $x$ that satisfy this 
condition, yielding a substantial numerical speedup for large atomic configurations.
In the following, for notational convenience, we suppress $\ell$ indices without loss 
of generality. Namely, we consider two local atomic environments $\tilde{x}$ and $\tilde{x}'$
centered on atomic positions $\mathbf{r}_\ell$ and $\mathbf{r}'_{\ell'}$.

The next step in building the SOAP kernel is to define a scalar product
between local atomic densities,
\begin{equation}
S(\tilde x,\tilde x')=\int_{\mathbb{R}^{3}}\rho_{\tilde x}(\mathbf{r})\rho_{\tilde x'}(\mathbf{r}) \, \mathrm{d}\mathbf{r}.\label{eq:S}
\end{equation}
The density fields $\rho_{\tilde x}$ and $\rho_{\tilde x'}$ are
naturally invariant to permutations of atomic indices, and $S$ inherits this symmetry.
Translation invariance is a consequence of comparing local environments.
However, $S$ is not invariant with respect
to rotations $\mathbf{r}\mapsto R\mathbf{r}$. To include this symmetry,
one may integrate over rotations,
\begin{equation}
\label{eq:haar_int}
k(\tilde x, \tilde x')=\int_{SO(3)} S(R\tilde x,\tilde x')^p \, \mathrm{d}R,
\end{equation}
where $\mathrm{d}R$ is the Haar measure and $R\tilde x$ generates
the configuration with density $\rho_{\tilde x}(R\mathbf{r})$. Numerical
evaluation of~\eqref{eq:haar_int} is non-obvious, and the SOAP approach involves
expanding $\rho_{\tilde x}(\mathbf{r})$ in terms of spherical harmonics
and applying orthogonality of the Wigner matrices.~\cite{Bartok2013repr}
In principle one could select integer $p>1$ as an arbitrary parameter, 
but in practice one typically fixes $p=2$ to simplify the expansion of spherical harmonics.

The final local SOAP kernel is defined by rescaling,
\begin{equation}
\label{eq:k_rescale}
\tilde{K}(\tilde x,\tilde x')=\left[ \frac{k(\tilde x,\tilde x')}
{\sqrt{k(\tilde x,\tilde x)k(\tilde x',\tilde x')}}\right]^{\zeta},
\end{equation}
for $\zeta >0$, which is possible because the unscaled kernel is strictly positive for nonempty environments.
After this
rescaling, we have $0\leq\tilde{K}(\tilde x,\tilde x')\leq1$ 
and $\tilde{K}(\tilde x,\tilde x)=1$. One commonly selects $\zeta>1$ to
effectively amplify the kernel in regions where $k(\tilde x,\tilde x')$
is largest.

By Eq.~\eqref{eq:kernel_identity}, the local kernel $\tilde K$ produces a total 
kernel $K$ that satisfies permutation, translation, and rotation invariance. The
hyperparameters are $\sigma$, $\Rc$, $p$, and $\zeta$.


\subsection{Random walk graph kernels}
\label{sec:graphtheory}

Here we present random walk graph kernels~\cite{gartner2002exponential, vishwanathan2010graph}.
In particular, the exponential graph kernel reviewed below will be
the basis of GRAPE, our method for kernel regression of potential energies (\emph{cf.} Sec.~\ref{sec:regression}). 
Our use of graphs to represent local atomic environments is inspired by
applications in chemical informatics~\cite{ralaivola2005graph,schietgat2008efficiently}.

A graph $G=(V,E)$ is defined as a set of vertices $V=(v_i)_{i=1\dots n}$, 
and edges $E\subseteq V\times V$ that connect pairs of vertices.
An unweighted graph is equivalently represented by its adjacency matrix $A\in\mathbb{R}^{n\times n}$,
where $A_{ij}=1$ if vertices $i$ and $j$ are connected by an edge
[\emph{i.e.} if $(v_{i},v_{j})\in E$] and $A_{ij}=0$ otherwise.

We work with weighted graphs, such that each edge $(v_{i},v_{j})\in E$
is assigned a nonzero weight $w_{ij}$. We represent such graphs via
the weighted adjacency matrix, 
\begin{equation}
\label{eq:adjacency}
A_{ij}=\begin{cases}
w_{ij} & \mbox{if } (v_{i},v_{j})\in E,\\
0 & \mathrm{otherwise}.
\end{cases}
\end{equation}
We consider undirected and unlabeled graphs, which means that $A_{ij}=A_{ji}$
and the nodes and edges have no additional structure.

Graph kernels measure similarity between two graphs. Many graph kernels
exist, including Laplacian kernels~\cite{smola2003kernels}, shortest
path kernels~\cite{borgwardt2005shortest}, skew spectrum kernels~\cite{kondor2008skew},
graphlet kernels~\cite{shervashidze2009efficient}, and functional
graph kernels~\cite{shrivastava2014new,shrivastava2014graph}.

Here we focus on random walk graph kernels because they are simple,
computationally efficient~\cite{vishwanathan2010graph}, and suitable for learning
potential energy landscapes, as will be made clear later.
The motivating idea is to consider a random walk over graph vertices
(\emph{i.e.} states). For this purpose, we assume for now  that $A$ represents the transition matrix
of a Markov process, such that $0\leq A_{ij}\leq1$ is the
probability for a random walker to transition from state $j$ to state $i$, and the columns of $A$ conserve probability, $\sum_{i=1}^{n}A_{ij} = 1$.
Later we will relax these probabilistic constraints and allow arbitrary $A_{ij} > 0$.
We select the initial state according to a distribution $p\in\R^{n}$,
\textit{i.e.} the walker starts at state $i$ with probability $0\leq p_{i}\leq1$.
Then $(A^{k}p)_{i}$ is the probability that the walker reaches
state $i$ after $k$ successful steps in the Markov chain. However, the walker may not reach $k$ steps. Analogous to $p$, we assume a stopping distribution $q\in\R^{n}$. That is, before each step, a walker in state $i$ stops with probability $0\leq q_{i}\leq1$. Then
\begin{equation}
\label{eq:factor_scalar}
s_{k}=\sum_{i=1}^{n}q_{i}(A^{k}p)_{i}=q^{T}A^{k}p
\end{equation}
represents the probability that a random walker stops after $k$ steps.

Our goal is to compare two graphs represented by adjacency matrices
$A\in\R^{n\times n}$ and $A'\in\R^{n'\times n'}$. For this, we consider simultaneous
random walks, one on each graph. The joint probability that the first
walker transitions from $j\rightarrow i$ and the second walker transitions
from $j'\rightarrow i'$ is the product of individual transition probabilities.
These joint probabilities appear as the elements of $A_{\times}\in\R^{nn'\times nn'}$, the direct
(Kronecker) product of adjacency matrices, defined by
\begin{equation}
A_{\times}=A\otimes A'=\begin{pmatrix}A_{1,1}A' & \hdots & A_{1,n}A'\\
\vdots &  & \vdots\\
A_{n,1}A' & \hdots & A_{n,n}A'
\end{pmatrix}.\label{eq:kronecker}
\end{equation}
Note that  that $A_{\times}$ can itself be interpreted
as an adjacency matrix for the so-called direct product graph~\cite{imrich2000product},
see Fig.~\ref{fig:productgraph}.

Given starting $p$, $p'$ and stopping $q$, $q'$ distributions, the probability that both walkers simultaneously stop after $k$ steps~\eqref{eq:factor_scalar} is
\begin{equation}
s_{k}s_{k}'=(q^{T}\otimes q'^{T})(A\otimes A')^{k}(p\otimes p')
=q_{\times}^{T}A_{\times}^{k}p_{\times},
\end{equation}
where $p_{\times}=p\otimes p'$, $q_{\times}=q\otimes q'$, and
we have used the matrix product identity $(A\otimes B)(C\otimes D)=(AC)\otimes(BD)$.

Two graphs can be compared by forming a weighted sum over paths of all lengths
of the Markov chain. The random walk graph kernel finally reads: 
\begin{equation}
\label{eq:graphkernel}
K(A,A')=\sum_{k\geq0} \mu_{k} s_{k}s_{k}'=q_{\times}^{T}\left(\sum_{k\geq0}
\mu_{k}A_{\times}^{k}\right)p_{\times}.
\end{equation}
The choice of $(\mu_k)_{k\geq0}$ is left to the user as long as the series
converges, and in this case~\eqref{eq:graphkernel} 
is known to define a positive semi-definite 
kernel~\cite{vishwanathan2010graph}.

One must also select the starting and stopping distributions. Choosing $p$, $p'$, $q$, $q'$ to be uniform guarantees invariance with respect to permutation
of vertices. To see this, consider permutation of the nodes, represented by
permutation matrices $T$ and $T'$,
which transform the adjacency matrices as $A\rightarrow TAT^{-1}$
and $A'\rightarrow T'A'T'^{-1}$. The product matrix transforms as
$A_{\times}\rightarrow(T\otimes T')A_{\times}(T\otimes T')^{-1}$,
where $(T\otimes T')^{-1}=(T^{-1}\otimes T'^{-1})$. Consequently,
the kernel~\eqref{eq:graphkernel} becomes $\hat{q}_{\times}^{T}\left(\sum_{\ell\geq0}
\mu_{\ell}A_{\times}^{\ell}\right)\hat{p}_{\times}$,
where $\hat{p}_{\times}=(T^{-1}p)\otimes(T'^{-1}p')$ and $\hat{q}_{\times}=(qT)\otimes(q'T')$.
If the starting and stopping distributions are uniform, we observe that 
$\hat{p}_{\times}=p_{\times}$ and $\hat{q}_{\times}=q_{\times}$, and the kernel 
exhibits permutation invariance.

Note, however, that with uniform starting and stopping distributions, it
is crucial that the columns of adjacency matrices $A$ and $A'$ \emph{not}
be normalized. If they were, then we would have $q^{T}A^{k}p=1$ 
independently of $A$, making it impossible to compare graphs. Consequently, 
although we continue to use~\eqref{eq:graphkernel}, we abandon its probabilistic interpretation.

In our work, we select $\mu_{k}=\gamma^{k}/k!$ in~\eqref{eq:graphkernel}
to get the exponential graph kernel~\cite{gartner2002exponential},
\begin{equation}
K(A,A')=q_{\times}^{T}\e^{\gamma A_{\times}}p_{\times},\label{eq:exponential}
\end{equation}
where $\gamma$ is a parameter controlling how fast the powers of $A_{\times}$ go to zero.
It can be shown to reweight the eigenvalues in the comparison 
process~\cite{vishwanathan2010graph,shrivastava2014new}.

The direct product matrix $A_{\times}$ contains $n^{2}n'^{2}$ matrix elements, a 
potentially large number, but the numerical cost of evaluating the kernel can
be reduced by diagonalizing the factor matrices.  Indeed, given $A=PDP^{-1}$ and $A'=P'D'P'^{-1}$,
with $D$ and $D'$ diagonal, we write
\begin{align}
A\otimes A' & =(PDP^{-1})\otimes(P'D'P'^{-1})\\
 & =(P\otimes P')(D\otimes D')(P\otimes P')^{-1},
\end{align}
so that~\eqref{eq:exponential} becomes 
\begin{equation}
K(A,A')=(q^{T}P\otimes q'^{T}P')\e^{\gamma D\otimes D'}(P^{-1}p\otimes P'^{-1}p').\label{eq:diagexponential}
\end{equation}
Because $D\otimes D'$ is itself a diagonal matrix, we can evaluate 
$\exp(\gamma D\otimes D')$ by applying
the exponential to each of the $nn'$ diagonal elements.

The computational cost to calculate $K(A,A')$ is dominated by matrix diagonalization, which scales like $\mathcal{O}(n^{3})$ assuming $n \sim n'$. If the eigen decompositions of $A$ and $A'$ have been precomputed, the scaling reduces to $\mathcal O(n^2)$.

\section{Graph Approximated Energy (GRAPE)}
\label{sec:grape}

The main contribution of our work is GRAPE, a random walk graph kernel tailored 
to local energy regression. Since random walk graph kernels act on the direct product of adjacency 
matrices, our primary task  is to select a suitable adjacency matrix 
associated with a local atomic environment $\tilde x$. To motivate our choice of 
adjacency matrix, we first present a graphical interpretation of the SOAP kernel.

The SOAP kernel is constructed from the inner product, Eq.~\eqref{eq:S}, between local atomic 
densities~\eqref{eq:local_density}. Combining these equations, we obtain a double 
sum over atoms from different local environments,
\begin{equation}
\label{eq:explicit}
\renewcommand{\arraystretch}{2} 
\begin{array}{ll} \displaystyle
S(\tilde x,\tilde x') &= \displaystyle
\sum_{i=1}^{\tilde n} \sum_{i'=1}^{\tilde n '} \omega(z_i) \omega(z_{i'})  \\
& \displaystyle \times f_{\mathrm{c}}(|\mathbf{r}_{i}|) f_{\mathrm{c}}(|\mathbf{r}'_{i'}|) 
\phi_{\sigma}(|\mathbf r_i - \mathbf r'_{i'}|), 
\end{array}
\end{equation}
where $\phi_{\sigma}$ is again Gaussian,
\begin{equation}
\label{eq:phidef}
\renewcommand{\arraystretch}{2} 
\begin{array}{ll} 
\displaystyle \phi_{\sigma}(|\mathbf r_i - \mathbf r'_{i'}|) &=  \displaystyle \int_{\mathbb R^3} \varphi_{\sigma} 
(|\mathbf r - \mathbf r_i|)   \varphi_{\sigma} (|\mathbf r - \mathbf r'_{i'}|)  \, d \mathbf r \\
&\displaystyle = \sigma^3 \pi^{3/2} \varphi_{\sqrt{2}\sigma} 
\left(|\mathbf r_i - \mathbf r'_{i'}|\right). 
\end{array}
\end{equation}
Here, for notational convenience, the positions of the atoms $\mathbf r_i$ and $\mathbf r'_{i'}$ are given relative to the centers $\mathbf r_\ell$ and $\mathbf r'_{\ell'}$ of the local environments.

\begin{figure}[!h]
\includegraphics[width=0.88\linewidth,trim = 0cm 10cm 0cm 0cm, clip]{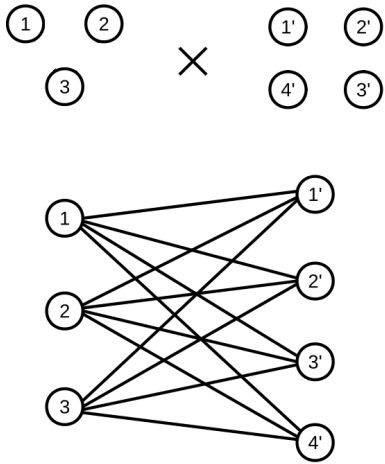}
\caption{Graph interpretation of the SOAP kernel. Two local environments comprised of 3 and 4 atoms are compared by forming a bipartite graph that links pairs of atoms.
}
\label{fig:crossgraph}
\end{figure}

Seeking a graph interpretation of~\eqref{eq:explicit}, we rewrite it in matrix form,
\begin{equation}
\label{eq:corrmatrix}
\left\{
\renewcommand{\arraystretch}{2} 
\begin{array}{rl} \displaystyle
S(\tilde x,\tilde x') &\displaystyle = \sum_{i=1}^{\tilde n} \sum_{i'=1}^{\tilde n'} 
B_{i i'}^2 = \mathrm{Tr}(B^T B), \\ \displaystyle
B_{i i'} &\displaystyle = \sqrt{\omega(z_i)\omega(z_i') f_{\mathrm{c}}(|\mathbf{r}_{i}|) 
f_{\mathrm{c}}(|\mathbf{r}'_{i'}|) \phi_{\sigma}( |\mathbf r_i - \mathbf r'_{i'}|)}. 
\end{array}
\right.
\end{equation}
The object $B\in\R^{\tilde n \times \tilde n'}$ is suggestive of an adjacency matrix. 
However, its indices $i$ and $i'$ reference atoms from different environments 
$\tilde x$ and $\tilde x'$. Indeed, the graphical interpretation of $B$ would be 
bipartite (\textit{i.e.}, containing only cross-links between the atoms of $\tilde x$ 
and $\tilde x'$) as illustrated in Fig.~\ref{fig:crossgraph}. Note that the 
elements $B_{i i'}$ vary with  the relative rotation between $\tilde x$ and $\tilde x'$,
and indeed SOAP requires nontrivial 
integration~\eqref{eq:haar_int} to achieve rotational invariance.

\begin{figure}[h]
\includegraphics[width=0.9\linewidth,trim = 0cm 15cm 0cm 0cm, clip]{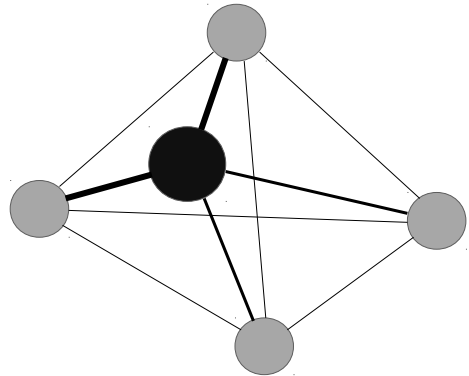}
\caption{Representation of methane $\mathrm{CH}_4$ as a graph.
Edge thickness between atoms $i$ and $j$ represents the adjacency matrix element $A_{ij}$.}
\label{fig:methanegraph}
\end{figure}

In GRAPE, we achieve rotational invariance by representing local configurations as graphs.
Guided by the form of~\eqref{eq:corrmatrix}, we introduce an analogous adjacency 
matrix $A\in\R^{\tilde n \times \tilde n}$ that operates on a \emph{single} local 
environment $\tilde x$,
\begin{equation}
\label{eq:adjconfig}
A_{ij} = \sqrt{\omega(z_i) \omega(z_j)   
f_{\mathrm{c}}(|\mathbf{r}_{i}|) f_{\mathrm{c}}(|\mathbf{r}_{j}|) 
\phi_{\sigma}(|\mathbf r_i - \mathbf r_j|)}.
\end{equation}
For example, Fig.~\ref{fig:methanegraph} illustrates the graph representation of a 
methane molecule. The elements $A_{ij}$ of the adjacency matrix~\eqref{eq:adjconfig} link atoms 
$i$ and $j$, both contained in $\tilde x$. Because these elements depend only on pairwise
distances, they are manifestly rotation invariant.

\begin{figure}[h]
\includegraphics[width=0.88\linewidth,trim = 0cm 10cm 0cm 0cm, clip]{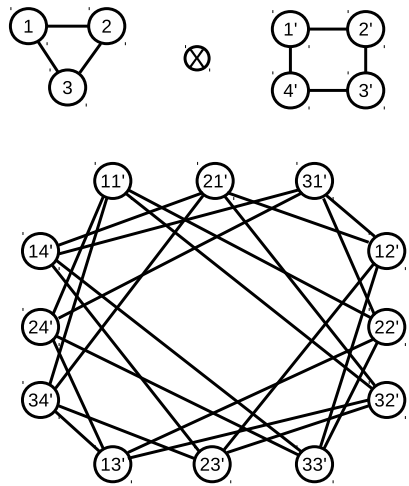}
\caption{Environments with $3$ and $4$ atoms are graphically represented by adjacency matrices $A$ and $A'$.
The direct product graph has $3 \times 4$ vertices and adjacency matrix elements
$(A \otimes A')_{ij,kl}=A_{ik}A'_{jl}$. Random walk graph kernels operate on this direct product matrix. }
\label{fig:productgraph}
\end{figure}

To compare two local atomic environments, we apply the exponential graph kernel 
of Sec.~\ref{sec:graphtheory}:
\begin{equation}
\label{eq:grape_k}
k(\tilde x, \tilde x')= (q^T \otimes q'^T) \e^{\gamma A \otimes A'} (p \otimes p').
\end{equation}
Figure~\ref{fig:productgraph} illustrates the graph represented by the direct 
product matrix, $A \otimes A'$, and should be contrasted with Fig.~\ref{fig:crossgraph}.
We again use uniform starting and stopping probability distributions 
$p$, $p'$, $q$, $q'$ to ensure invariance with respect to permutation of indices, while
rotational invariance is naturally inherited from the adjacency matrices $A$ and $A'$.

Finally, we use~\eqref{eq:k_rescale} to rescale~\eqref{eq:grape_k}, 
and~\eqref{eq:local_kernel} to build our model which is suitable for regression on 
total energies. The kernel is regular and its derivatives are calculable in closed form 
(Appendix~\ref{sec:forces}), so atomic forces are also available from the regression model.

GRAPE shares most of its hyperparameters with SOAP, 
namely: $\sigma$, $\sigma_{\mathrm{cut}}$, and $\zeta$. However, the SOAP 
hyperparameter $p$ is replaced by GRAPE's $\gamma$.
Note that $\gamma$ enables effective reweighting of the eigenvalues, each of which corresponds to a different length scale~\cite{vishwanathan2010graph,shrivastava2014new,shrivastava2014graph}.
For all methods considered in this work, we must also specify the
regularization parameter $\lambda$ used in kernel ridge regression~\eqref{eq:local_kernel}.

The computational cost to directly evaluate the local GRAPE kernel $k(\tilde x, \tilde x')$ scales 
as $C_\mathrm{loc} \sim \tilde n^3$ (see Sec.~\ref{sec:graphtheory}), where 
$\tilde n$ is the typical number of atoms in local environments $\tilde x$ 
and $\tilde x'$. Global energy 
regression~\eqref{eq:local_kernel} requires many local kernel evaluations, for both SOAP 
and GRAPE methods. Evaluating the total kernel~\eqref{eq:kernel_identity} requires 
double summation over all $n \sim n'$ atoms in total configurations $x$ and $x'$, and 
thus scales as $\mathcal O(n^2 C_\mathrm{loc})$. If there are $N$ atomic 
configurations $(x_i)_{i=1\dots N}$ in the dataset, then 
evaluating the full kernel matrix $K$ scales as $\mathcal O(N^2 n^2 C_\mathrm{loc})$. 
Matrix inversion brings the total scaling to 
$\mathcal O(N^2 n^2 C_\mathrm{loc} + N^3)$. An improvement is to
precompute the diagonalization of every local adjacency matrix at cost 
$\mathcal O(N n \tilde n^3)$. Then subsequent local kernel evaluations scale 
like $C_\mathrm{loc} \sim \tilde n^2$ and the total cost of building the GRAPE energy 
regression model becomes $\mathcal O(N n \tilde n^3 + N^2 n^2 \tilde n^2 + N^3)$.
Once the model is built, computing the approximation energy~\eqref{eq:local_kernel} for a new configuration
scales as $\mathcal{O}( n \tilde{n}^3 + Nn^2\tilde{n}^2)$.


\section{Benchmark}
\label{sec:application}

We demonstrate the competitiveness of GRAPE by benchmarking it on a standard 
energy regression problem. We use the QM7 dataset of organic models used 
in Ref.~\onlinecite{Rupp2012} and freely available at 
\url{http://quantum-machine.org/datasets/}. This database contains $7165$ molecules randomly 
selected from the GDB-13 database with associated atomization energies, typically between 
$-2000$ and~$-800$ kcal/mol, obtained from hybrid DFT 
calculations~\cite{PhysRev.136.B864,PhysRev.140.A1133}. GDB-13 contains approximately 
$10^9$ organic molecules up to $23$ atoms in size, and formed from elements H, C, N, O, and S.
QM7 has been randomly partitioned into $5$ sequences, each containing 1433 molecules.
We use the first $N$ molecules of Partition 1 as training data 
(with $N = 100, 300, 500, 1,000$), and Partitions 2 and 3 as validation data, 
\textit{i.e.} to select the hyperparameters necessary for the various regression 
methods (discussed below). Finally, after having locked the hyperparameters, we 
use Partition 4 as our test data, from which we estimate the mean absolute error 
(MAE) and root mean square error (RMSE) of the energy regression models.

We compare the performance of GRAPE against that of the Coulomb
matrix method~\cite{Rupp2012} (Appendix~\ref{sec:coulomb}) and SOAP~\cite{bartok2015gaussian}.
After experimenting on the validation data, we selected the following hyperparameters for 
these methods. We set the Gaussian width parameter in~\eqref{eq:soapphi} to be 
$\sigma=1.0\,\angstrom$ for both SOAP and GRAPE. This provides a small overlap of the 
Gaussian densities of bonded atoms, which are typically separated by a few $\angstrom$. We 
select the cutoff radius in~\eqref{eq:soapfc} to be $\Rc = 4.0 
\,\angstrom$ for both SOAP and GRAPE, such that a typical local atomic neighborhood contains 
around 5 atoms. Interestingly, we find that GRAPE is especially sensitive to this 
hyperparameter, and that the energy regression error would nearly double if we instead 
selected $\Rc = 5.0\,\angstrom$ for GRAPE.
We must also select the weight $\omega(z_i)$ in~\eqref{eq:local_density} 
and~\eqref{eq:adjconfig} as a function of atomic number $z_i$. Again, after some 
experimentation, we select $\omega(z_i) = z_i$ for SOAP and $\omega(z_i) = 1/z_i$ for GRAPE.
The latter implies that the GRAPE hyperparameter $\gamma$ appearing in~\eqref{eq:grape_k} 
should scale  like the square of the typical atomic number in the data; we 
select $\gamma = 40.0$. Consistent with previous work~\cite{Bartok2013repr}, we select 
$p=2$ for SOAP~\eqref{eq:haar_int}  and $\zeta = 4.0$ in~\eqref{eq:k_rescale} for 
both SOAP and GRAPE. We select $\alpha = 0.05$ for the Coulomb hyperparameter appearing in 
Eq.~\eqref{eq:coul_kernel}. The last hyperparameter $\lambda$, appearing in~\eqref{eq:local_kernel}, regularizes the kernel regression. We select $\lambda = 10^{-7}$ for 
Coulomb, $\lambda = 10^{-5}$ for SOAP, and $\lambda = 10^{-3}$ for GRAPE.  
In selecting the above hyperparameters, we put 
approximately equal weight on the MAE and RMSE. We caution that all three models (Coulomb, 
SOAP, GRAPE) would likely benefit from a more exhaustive search over the hyperparameters.
Our results should thus be interpreted as a \emph{qualitative} comparison of GRAPE's 
performance relative to Coulomb and SOAP.

\begin{table}
\centering
\newcolumntype{C}[1]{>{\centering\let\newline\\\arraybackslash\hspace{0pt}}m{#1}}
\renewcommand{\arraystretch}{1.5} 
\begin{tabular}{C{1.4cm}|C{1.2cm}||C{1.3cm}C{1.3cm}C{1.3cm}C{1.3cm}}
& $N$ & 100 & 300 & 500 & 1,000 \\ 
\hline \hline
\multirow{2}{*}{Coulomb} 
& MAE & 25.6 & 19.8 & 17.9 & 17.7  \\ 
& RMSE & 50.8 & 33.5 & 27.1 & 28.5  \\  \hline
\multirow{2}{*}{SOAP} 
& MAE & 15.6 & 11.3 & 10.4 & 9.7  \\ 
& RMSE & 21.0 &  15.6 & 14.5 & 13.3  \\  \hline
\multirow{2}{*}{GRAPE} 
& MAE & 11.2 & 10.1 & 9.6 & 9.0  \\ 
& RMSE & 14.9 & 13.9 & 13.3 &  12.7 \\  \hline
\end{tabular}
\caption{Mean Absolute Error (MAE) and Root Mean Square Error (RMSE), in units of kcal/mol, for Coulomb, SOAP and GRAPE regression methods, and various sizes $N$ of the training dataset.}
\label{tab:errors_regression}
\end{table}

\begin{figure*}[ht]
\includegraphics[width=1.0 \linewidth]{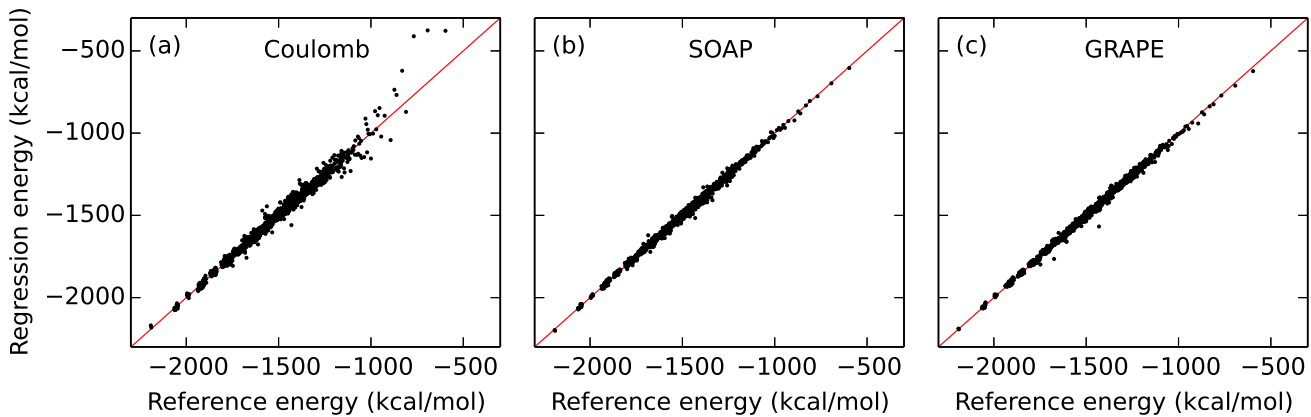}
\caption{Atomization energies predicted by (a) Coulomb, (b) SOAP, and (c) GRAPE methods. The three regression models were built using a training dataset of $N=1,000$ organic molecules. Reference energies were calculated using density functional theory.}
\label{fig:energy}
\end{figure*}

Table~\ref{tab:errors_regression} displays the MAE and RMSE error estimates for all three
methods and various sizes $N$ of the training dataset. Figure~\ref{fig:energy} shows 
regression energies for the test data with $N=1,000$, and we observe that GRAPE is competitive 
with SOAP; given our limited tuning of the hyperparameters, we cannot conclude that one 
method outperforms the other. However, both SOAP and GRAPE significantly and consistently 
outperform the Coulomb method. We note that our MAE and RMSE estimates for the Coulomb 
method are consistent with those previously reported~\cite{Rupp2012}, when holding $N$ 
fixed.

For reference, the best known methods in the literature outperform all three kernels in Table~\ref{tab:errors_regression}.
Reference~\onlinecite{Molecular2013} achieves an MAE of $\approx 6$ kcal/mol for $N=1,000$ using a Laplacian kernel applied to a random sorting of the Coulomb matrix.~\cite{Montavon2012}
The Bag-of-Bonds kernel method~\cite{Hansen2015} reduces the MAE to $\approx 3.3$ for $N=1,000$.
Neural networks appear to be state of the art for this QM7 dataset. When training on $N=5,800$ molecules, the Deep Tensor Neural Network~\cite{Schutt2017} achieves an MAE of $\approx 1.0$ kcal/mol---a significant improvement over the Bag-of-Bonds MAE, $\approx 1.5$ kcal/mol.

\section{Discussion}
\label{sec:discussion}
We introduced GRAPE, a simple and flexible method for energy regression based on 
random walk graph kernels. The approach is invariant with respect to translations,
rotations, and permutations of same-species atoms.
Using a standard benchmark dataset 
of organic molecules, we demonstrated that GRAPE's energy predictions are competitive with commonly applied 
kernel methods. 
Because the GRAPE kernel is regular and admits simple closed-form derivatives, it is a candidate for
fitting atomic forces, although we did not benchmark the method on this application.

Like the Coulomb method~\cite{Rupp2012}, GRAPE essentially consists of comparing matrices.
These matrices, built upon interatomic distances, are inherently invariant with respect 
to translation and rotation, while carrying complete information of the system. However, 
the matrix representation lacks permutation invariance. To restore permutation invariance, 
the Coulomb method restricts its attention to the list of ordered eigenvalues.
Unfortunately, due to this sorting, the associated kernel becomes
non-differentiable, which can be a source of instability, as pointed out 
in Ref.~\onlinecite{hirn2016wavelet}. Unlike Coulomb, GRAPE is completely regular.

An alternative machine learning approach is to begin by representing 
the atomic configuration as a density, \textit{e.g.} using Gaussians peaked on 
each atomic position. This density field provides a natural way to compare two environments~\cite{Bartok2010GAP,ferre2015permutation,hirn2015quantum},
and this is the approach taken by the SOAP method~\cite{Bartok2013repr}.
A disadvantage of this approach is that the density is not inherently rotational invariant, and density-based kernels typically require an explicit integration over rotations.
An achievement of the SOAP kernel is that its rotational integral, Eq.~\eqref{eq:haar_int},
can be evaluated in closed form via a spherical harmonic expansion of the Gaussian densities.
GRAPE obviates the need for such an integral in the first place.

An exciting aspect of the GRAPE approach is its flexibility for future extensions.
For example, one can use multiscale analysis techniques, such as the multiscale Laplacian graph 
kernel~\cite{kondor2016multiscale}, or wavelet and scattering transforms over 
graphs~\cite{smalter2009graph,hammond2011wavelets,chen2014unsupervised}. 
Another promising research direction is to encode more physical knowledge into the
graph, \textit{e.g.} using labeled graphs~\cite{vishwanathan2010graph}. The vertex labels could directly represent  the atomic species, or could more subtly encode chemical information such 
as the number of electrons in the valence shell.

\acknowledgments
The authors thank G\'{a}bor Cs\'{a}nyi, Danny Perez,
Arthur Voter, Sami Siraj-Dine and Gabriel Stoltz for fruitful
discussions. Work performed at LANL was supported by the Laboratory Directed Research and Development (LDRD) and the Advanced Simulation and Computing (ASC) programs. 

\appendix

\section{Kernel ridge regression}
\label{sec:kernelapprox}

Here we review the machine learning technique of kernel ridge regression.
We assume a dataset of pairs $(x_{i},y_{i})_{i=1\dots N}$. In
our application, each $x_{i}$ represents an atomic configuration
and $y_{i}$ its corresponding energy. The goal is to build a model
of energies $\hat{E}(x)$ for configurations $x$ not contained in the dataset.
Following Ref.~\onlinecite{Hastie2009_krr} we begin with ordinary
ridge regression.

We formally model the energy as a linear combination of 
descriptors $(h_{m})_{m=1\dots M}$,
\begin{equation}
\hat{E}(x)=\sum_{m=1}^{M}\beta_{m}h_{m}(x),\label{eq:f_decomp_app}
\end{equation}
where $M$ is possibly infinite. Note that $\hat{E}(x)$ is linear in the
regression coefficients $\beta_{m}$ but potentially very nonlinear
in $x$. We select regression coefficients $\hat{\beta}_{m}$
that minimize a loss function
\begin{equation}
\sum_{i=1}^{N}V(y_{i}-\hat{E}(x_{i}))+\frac{\lambda}{2}\sum_{m=1}^{M}\beta_{m}^{2}.\label{eq:ridge_cost}
\end{equation}
The simplest error measure, $V(r)=r^{2}$, yields a linear system of equations 
to be solved for $\beta_m$. The solution is the well known ridge
regression model~\cite{Hoerl1970}:
\begin{align}
\hat{E}(x) & =\sum_{m=1}^{M}\hat{\beta}_{m}h_{m}(x),\label{eq:rr_1_app}\\
\hat{\beta} & =(HH^{T}+\lambda I)^{-1}H  y,\label{eq:rr_2_app}
\end{align}
with matrix elements $H_{mi}=h_{m}(x_{i})$ and vector $ y = (y_1, y_2, \dots, y_N)$. 
The empirical regularization parameter $\lambda>0$ effectively smooths the approximation 
$\hat{E}$, and consequently improves the conditioning of the 
linear inversion problem.

It turns out that we can generalize this model by introducing the inner product kernel,
\begin{equation}
K(x,x')=\sum_{m=1}^{M}h_{m}(x)h_{m}(x'),\label{eq:kernel_app}
\end{equation}
which measures similarity between inputs $x$ and $x'$. Note that
\begin{align}
K_{ij}=K(x_{i},x_{j}) & =(H^{T}H)_{i,j}, \\
K(x,x_{i}) & =( h^{T}H)_{i},
\end{align}
where $h = (h_1(x), h_2(x), \dots, h_M(x))$. The matrix identity 
$(HH^{T}+\lambda I)^{-1}H = H (H^{T} H+\lambda I)^{-1}$ allows us to 
rewrite~(\ref{eq:rr_1_app}) and~(\ref{eq:rr_2_app}) in the suggestive form:
\begin{align}
\hat{E}(x) & =\sum_{i=1}^{N}\hat{\alpha}_{i}K(x,x_{i}),\label{eq:krr_1_app}\\
\hat{\alpha} & =(K+\lambda I)^{-1}  y.\label{eq:krr_2_app}
\end{align}
The key insight is that the descriptors $h_{m}$ no longer appear
explicitly. One may select the kernel $K$ directly, and thus
implicitly define the family $(h_{m})_{m=1 \dots M}$.

Equations~\eqref{eq:krr_1_app} and~\eqref{eq:krr_2_app} together with
a suitable kernel constitute the method of kernel ridge regression.
The kernel must be symmetric and positive, \textit{i.e.} $K(x,x')=K(x',x)$ for all $x$, $x'$,
and $\sum_{i}\sum_{i'}\alpha_{i}\alpha_{i'}K(x_{i},x_{i'})\geq0$
for any non-empty collection of points and coefficients $\{(x_{i},\alpha_{i})\}$.
By Mercer's theorem~\cite{Mercer1909}, these conditions are equivalent
to the decomposition~\eqref{eq:f_decomp_app}.

There are multiple ways to interpret Eqs.~\eqref{eq:krr_1_app} and~\eqref{eq:krr_2_app}.
In statistical learning theory, we may derive $\hat{E}(x)$ as the function
in some space $\mathcal{H}$ 
that best minimizes the squared error 
$\sum_{i=1}^N\left(y_{i}-\hat E(x_{i})\right)^{2}$ subject 
to a regularization term $\lambda||\hat{E}||_{\mathcal{H}}$. The kernel $K$ generates
both $\mathcal{H}$ (the so-called reproducing kernel Hilbert space)
and its norm $||\cdot||_{\mathcal{H}}$~\cite{Smola1998,scholkopf2001learning}.
Another interpretation is Bayesian. In this case, one views the coefficients $\beta_{m}$
as independent random variables with Gaussian prior probabilities; then the
kernel $K$ specifies the covariance of $y$ and $y'$, and
$\hat{E}(x)$ becomes the posterior expectation for configuration $x$. This approach is called
Gaussian process regression or Kriging~\cite{Rasmussen2006}.

We note that Eq.~\eqref{eq:krr_1_app} [but not \eqref{eq:krr_2_app}] is
independent of the specific choice of error measure $V$. For example,
the method of support vector regression corresponds to the choice
$V(r)=\max(0,|r|-\epsilon)$.


\section{Local energy decomposition}
\label{sec:local}

We assume that the energy of an atomic configuration
$x$ with $n$ atoms can be decomposed as a sum over local atomic environments $\tilde x_\ell$,
\begin{equation}
E(x)=\sum_{\ell=1}^{n}\varepsilon(\tilde x_\ell).\label{eq:locality_app}
\end{equation}

Following Eq.~(\ref{eq:f_decomp_app}), we model local energies $\varepsilon(\tilde x_\ell)$
as a linear combination of abstract descriptors $\tilde{h}_{m}(\tilde x_\ell)$,
\begin{equation}
\hat{\varepsilon}(\tilde x_\ell)=\sum_{m=1}^{M}\beta_{m}\tilde{h}_{m}(\tilde x_\ell).\label{eq:local_model}
\end{equation}
After inserting~\eqref{eq:local_model} into~\eqref{eq:locality_app}, we observe 
that the model for total energy
\begin{equation}
\hat{E}(x)=\sum_{\ell=1}^{n}\hat{\varepsilon}(\tilde x_\ell)=\sum_{m=1}^M \beta_{m}h_{m}(x),
\end{equation}
involves the same coefficients $\beta_{m}$ but new descriptors 
\begin{equation}
h_{m}(x)=\sum_{\ell=1}^{n}\tilde{h}_{m}(\tilde x_\ell).\label{eq:global_h}
\end{equation}

We assume a dataset $(x_{i},y_{i})_{i=1\dots N}$ containing configurations
$x_{i}$ and total energies 
\begin{equation}
\label{eq:total_energy_y}
y_{i}=E(x_{i}).
\end{equation}
Again minimizing the cost function of Eq.~(\ref{eq:ridge_cost}),
the kernel ridge regression model of Eqs.~(\ref{eq:krr_1_app}) and~(\ref{eq:krr_2_app})
is unchanged. However, the kernel between the full atomic configurations
is now constrained. Equations~(\ref{eq:kernel_app}) and~(\ref{eq:global_h})
together imply,
\begin{equation}
K(x,x')=\sum_{\ell=1}^{n}\sum_{\ell'=1}^{n'}\tilde{K}(\tilde x_\ell,
\tilde x'_{\ell'}).\label{eq:kernel_identity_app}
\end{equation}
where $\tilde{K}(\tilde x_\ell,\tilde x'_{\ell'})=\sum_{m}\tilde{h}_{m}(\tilde x_\ell)\tilde{h}_{m}(\tilde x'_{\ell '}).$

We conclude that the energy decomposition~(\ref{eq:locality_app}) constrains 
the kernel $K$ to a sum over terms $\tilde{K}$ involving \emph{local} environments. 
In ``kernelizing'' this model, we only require specification
of the local kernel $\tilde{K}$; the descriptors $\tilde{h}_{m}$
are implicit and typically infinite in number.

\section{Coulomb matrix method}
\label{sec:coulomb}

We review the Coulomb matrix method of Ref.~\onlinecite{Rupp2012}. Here,
we elide localized kernel~\eqref{eq:local_kernel} and consider only global atomic 
configurations $x$ as in Ref.~\onlinecite{Rupp2012}. Given a configuration with 
positions and atomic numbers $(\mathbf{r}_{i},z_i)_{i=1\dots N}$, the Coulomb matrix is
\begin{equation}
\label{eq:coulomb}
M_{i,j}=\left\{ \global\long\def\arraystretch{1.5}
\begin{array}{cc}
\frac{z_{i}z_{j}}{\|r_{i}-r_{j}\|}, & i\neq j,\\
0.5 \, z_{i}^{2.4}, & i=j.
\end{array}\right..
\end{equation}
Although the matrix is invariant with respect to translations and
rotations of the atomic positions, it is not invariant with respect
to permutations of indices. The matrix eigenvalues, however, are permutation
invariant. To compare configurations $x$ and $x'$ (of size $n$
and $n'$, respectively), the Coulomb method compares the lists of eigenvalues
($\lambda_{i}$ and $\lambda'_{i}$, respectively), sorted by decreasing magnitude and truncated at length
$\min(n,n')$. We define normalized eigenvalues,
\begin{align}
\hat{\lambda}_{i} & =\text{\ensuremath{\lambda}}_{i}/||\lambda||,\\
\hat{\lambda}'_{i} & =\text{\ensuremath{\lambda}}'_{i}/||\lambda'||,
\end{align}
where $||\lambda||=\sqrt{\sum_{i=1}^{n}\lambda_{i}^{2}}$. Then the Coulomb kernel reads
\begin{equation}
\label{eq:coul_kernel}
K(x,x')=\exp\left[-\alpha \, \sum_{i=1}^{\min(n,n')}(\hat{\lambda}_{i}-\hat{\lambda}'_{i})^{2}\right],\end{equation}
with $\alpha$ a hyperparameter.

\section{Computing forces}
\label{sec:forces}

Given a regression model for energy, it is often desirable to compute its gradient 
to obtain a regression model for forces.
One motivation is to use machine learned forces within molecular dynamics
simulations. Another motivation is that typical datasets (\textit{e.g.} as
generated by \emph{ab initio }calculations) often contain force information
that can aid the energy regression~\cite{bartok2015gaussian}.
In either case, the key step is to calculate the derivative $\partial_{\mathbf r} K(x,x')$
of the kernel $K(x,x')$ with respect to some atomic position $\mathbf r$ in configuration $x$.
Here, we consider the exponential graph kernel~\eqref{eq:grape_k} used in GRAPE.
As before, $A$ and $A'$ denote the adjacency matrices of $x$ and
$x'$ respectively. Assuming $\partial_{\mathbf r} A' = 0$, the gradient of the unscaled kernel is,
\begin{equation}
\label{eq:kernelderiv}
\partial_{\mathbf r} k(x,x')=\gamma(q^{T}\otimes q'^{T})(\partial_{\mathbf r} A \otimes A')
e^{\gamma A \otimes A'}
(p \otimes p').
\end{equation}

Referring to Eq.~\eqref{eq:adjconfig} we see that the GRAPE adjacency matrix element
$A_{ij}$ has a simple functional dependence on positions $\mathbf{r}_{i}$
and $\mathbf{r}_{j}$ (and also an implicit dependence on the central
atom position $\mathbf{r}_{\ell}$), and it is thus straightforward to express $\partial_{\mathbf r} A$
in closed form. To compute the normalized kernel~\eqref{eq:k_rescale}, 
one also needs:
\begin{equation}
\label{eq:kernelderiv}
\partial_{\mathbf r} k(x,x)=2\gamma(q^{T}\otimes q^{T})(\partial_{\mathbf r} A \otimes A)
e^{\gamma A \otimes A}
(p \otimes p).
\end{equation}

\bibliographystyle{apsrev4-1.bst}
%

\end{document}